\documentclass[prb,preprintnumbers,superscriptaddress,amsmath,amssymb,floatfix,aps,twocolumn]{revtex4}
\usepackage{graphicx}
\usepackage{lipsum} 
\usepackage{comment}
\usepackage{lineno}

\usepackage{xcolor}

\usepackage{ulem}

\usepackage{afterpage}

\usepackage[version=4]{mhchem}

\usepackage{hyperref}
\usepackage{amsmath}
\usepackage{amsbsy}
\usepackage{float}
\usepackage[utf8]{inputenc}

\begin{document}

\title{Phase engineering of MoS$_2$ monolayers: A pathway to enhanced lithium-polysulfide battery performance}

\date{\today} 

\author{J. W. Gonz\'alez}
\email{jhon.gonzalez@uatonf.cl}
\affiliation{Departamento de Física, Universidad de Antofagasta, Av. Angamos 601, Casilla 170, Antofagasta, Chile}

\author{E. Flórez}
\affiliation{Instituto  de Ciencias  Básicas, Universidad de  Medellín, Medellín-Colombia}

\author{R. A. Gallardo}
\affiliation{Departamento de Física, Universidad Técnica Federico Santa María, Avenida España 1680, Valparaíso, Chile}

\author{J. D. Correa}
\affiliation{Instituto  de Ciencias  Básicas, Universidad de  Medellín, Medellín-Colombia}

\begin{abstract}
This study explores the potential of MoS$_2$ polymorphs, specifically the semiconducting 2H phase and the metallic 1T$^\prime$ phase, as anchoring materials to enhance the electrochemical performance of lithium-sulfur (Li--S) batteries. Using density functional theory calculations, we show that 1T$^\prime$-MoS$_2$ exhibits stronger Li--S interactions, greater charge transfer, and enhanced catalytic activity compared to its 2H counterpart, effectively suppressing polysulfide dissolution and facilitating redox reactions. The reversible 2H$\leftrightarrow$1T$^\prime$ transition offers a tunable design space for balancing conductivity and structural stability. These findings position hybrid MoS$_2$ architectures as promising platforms for next-generation Li--S batteries with improved energy density, cycling stability, and rate capability.
\end{abstract}

\maketitle

\section{\label{sec:intro} Introduction}

The global transition to low-carbon energy systems demands high-performance energy storage technologies to support electric transportation, grid-scale systems, and portable electronics.\cite{gielen2019role,mahlia2014review,kalair2021role} 
Lithium–polysulfur (Li–S) batteries are among the most promising alternatives to conventional lithium-ion batteries (LIBs), offering a theoretical energy density of approximately 2600~Wh\,kg$^{-1}$ and a specific capacity of 1675~mAh\,g$^{-1}$, values two to three times higher than those of LIBs.\cite{Ji2009,Li2018} Sulfur is also abundant, low-cost, and environmentally benign, making Li–S chemistry attractive for next-generation applications.\cite{manthiram2020reflection,su2012lithium}

\begin{figure*}[!]
\centering
\includegraphics[clip,width=0.76\textwidth,angle=0]{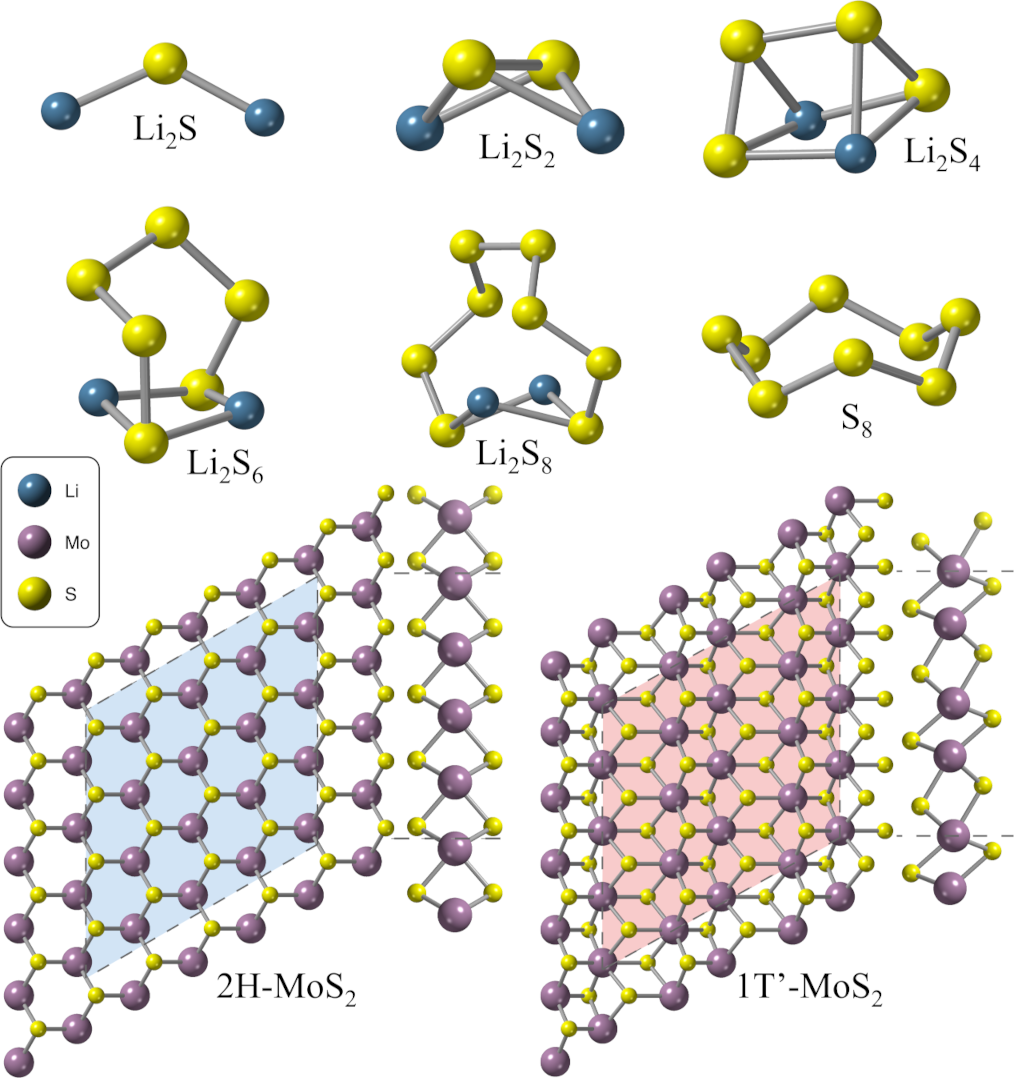} 
\caption{Schematic representation of various lithium sulfide (Li$_m$S$_n$) clusters and monolayer MoS$_2$ structures. The top panel illustrates the geometries of the Li$_m$S$_n$ molecules studied, while the bottom panel depicts the 2H-MoS$_2$ (left) and 1T$^\prime$-MoS$_2$ (right) phases. For the MoS$_2$ monolayers, both top (view along the c-axis) and side (view along the b-axis) perspectives are presented. The shaded regions in the bottom panel indicate the supercells under investigation. Lithium (Li) is shown in blue, molybdenum (Mo) in purple, and sulfur (S) in yellow. 
}
\label{Fig:scheme}
\end{figure*}

Li--S batteries operate through a multistep redox mechanism in which elemental sulfur (S$_8$) is progressively reduced to form lithium polysulfides (Li$_2$S$_n$), including long-chain species (Li$_2$S$_6$, Li$_2$S$_8$) and short-chain intermediates (Li$_2$S$_2$, Li$_2$S$_4$), before ultimately yielding solid Li$_2$S during discharge.\cite{Su2012} This sequence governs critical aspects of cell performance, such as Coulombic efficiency, self-discharge, and cycle life. Lithium polysulfides (Li$_m$S$_n$) thus play a central role in determining the electrochemical behavior of Li--S cells. Figure~\ref{Fig:scheme}(top) shows the DFT-optimized geometries of key intermediates. Modeling the full polysulfide manifold is essential to reproduce the shuttle effect and to evaluate surface-engineering strategies that suppress it;\cite{li2018revisiting,song2020rationalizing} in this work, we address this challenge by focusing on a representative species (Li$_2$S$_4$), which captures key features of the intermediate redox steps.

Despite their theoretical advantages, Li–S batteries face three persistent challenges. First, long-chain Li$_2$S$_n$ species are soluble in typical electrolytes and tend to diffuse toward the anode, causing self-discharge and capacity fading. Second, sulfur and its reduced forms exhibit extremely low electronic conductivity (e.g., $<10^{-30}$~S\,cm$^{-1}$ for S$_8$), which hinders charge-transfer kinetics. Third, the large volumetric expansion (up to 80\%) during full lithiation compromises structural integrity and limits cyclability.\cite{Song2020}

To address these limitations, two-dimensional (2D) materials have been proposed as functional hosts that can immobilize polysulfides, catalyze redox reactions, and buffer volume changes.\cite{shao2019two,glavin2020emerging,khossossi2020rational,fan2022two} Among these, molybdenum disulfide (MoS$_2$) stands out for its rich polymorphism and tunable properties. This work focuses on the semiconducting 2H phase and the metallic 1T$^\prime$ phase, which exhibit complementary functionalities as cathode materials.\cite{voiry2015phase,Wang2019,ni2024perspectives} The 2H phase can suppress parasitic reactions due to its low conductivity, while the 1T$^\prime$ phase promotes redox reactions through enhanced charge transfer and catalytic activity.
Importantly, the 2H $\leftrightarrow$ 1T$^\prime$ phase transition is reversible and can be induced via chemical doping, strain, or thermal treatment.\cite{Zheng2021,lukowski2013enhanced} This tunability enables the design of hybrid architectures that combine the stability of semiconducting regions with the electrochemical reactivity of metallic ones, enhancing both rate performance and cycling stability.\cite{Li2017,zhao2023recent}

In this work, we use density functional theory (DFT) to investigate the interaction between Li$_m$S$_n$ clusters and MoS$_2$ monolayers in 2H and 1T$^\prime$ phases. We compute adsorption energies, charge transfer, structural rearrangements, and delithiation pathways across the polysulfide series. By comparing how each polymorph stabilizes short- and long-chain intermediates, we uncover phase-engineering principles that enable effective polysulfide immobilization without compromising redox kinetics. These insights offer atomistic-level guidance for the design of high-performance Li–S battery cathodes based on tunable 2D materials.

\section{\label{sec:comp} Computational Details}

Density functional theory (DFT) calculations are performed using the Vienna Ab-initio Simulation Package (VASP).\cite{VASP0} Exchange–correlation effects are described within the generalized gradient approximation (GGA) using the Perdew–Burke–Ernzerhof (PBE) functional,\cite{Perdew2008} including long-range van der Waals interactions via the DFT-D3 method with Becke–Johnson damping.\cite{grimme2010consistent,grimme2011effect} The projector-augmented wave (PAW) method is employed, using the standard pseudopotentials recommended by VASP. The valence electron configurations are: Mo ($4s^2\,4p^6\,4d^5\,5s^1$), S ($3s^2\,3p^4$), and Li ($1s^2\,2s^1$). A plane-wave energy cutoff of 450~eV ensures total energy convergence.
The Brillouin zone is sampled using a $\Gamma$-centered Monkhorst–Pack grid of $5\times5\times1$, yielding a reciprocal-space resolution of approximately $0.02\times2\pi$/\AA{}. All atomic positions are relaxed until residual forces fall below $10^{-3}$~eV/\AA. Total energies, electronic structures, and adsorption properties are obtained from fully relaxed geometries.

To calculate thermodynamic quantities, we complement periodic DFT with gas-phase quantum chemical calculations using the ORCA package (version 5.0.4).\cite{Neese2012,Neese2020} The M06-L meta-GGA functional\cite{Zhao2008} is used with the def2-TZVP basis set, along with the def2/J auxiliary basis for Coulomb fitting. All simulations employ a fine numerical integration grid (DEFGRID3) and tight SCF convergence criteria with an energy threshold of $1.0 \times 10^{-8}$~a.u.
Vibrational frequencies are computed within the harmonic approximation to obtain zero-point energies and thermal corrections to the enthalpy and Gibbs free energy. This molecular approach efficiently evaluates charge transfer, vibrational modes, and thermodynamic properties without the complexity of periodic boundary conditions.

\section{\label{sec:results} Results and Discussion}

Herein presents a comprehensive first-principles analysis of lithium polysulfide interactions with MoS$_2$ monolayers in both 2H and 1T$^\prime$ phases. We begin by characterizing the adsorption geometries and energetics across the Li$_m$S$_n$ series, then by evaluating charge transfer, electronic structure modifications, and delithiation pathways. Finally, we incorporate vibrational and solvation effects in a gas-phase model to assess the thermodynamic viability of adsorption under realistic battery operating conditions.

\subsection{Pristine MoS$_2$}

We consider two metastable polymorphs of molybdenum disulfide (MoS$_2$): the semiconducting 2H phase with a trigonal prismatic structure (space group $P6_3/mmc$) and the metallic 1T$^\prime$ phase with a distorted octahedral geometry (space group $C2/m$).\cite{voiry2015phase,krishnan2019synoptic} Figure~\ref{Fig:scheme}(bottom) shows the atomic structures of both phases. We construct supercells containing 16 Mo and 32 S atoms for each case to enable direct comparison.

For 2H-MoS$_2$, we use a $4\times4\times1$ supercell with lattice constants $a = b = 12.61$~\AA{} and a layer thickness of approximately 3.14~\AA. The Mo-S and Mo-Mo bond lengths are 2.40~\AA{} and 3.15~\AA{}, respectively, in agreement with experimental and computational data.\cite{fu2013first} In contrast, the 1T$^\prime$ phase is modeled using a $2\times4\times1$ supercell with $a = 12.96$~\AA{} and $b = 12.59$~\AA{}, and a layer thickness of 3.47~\AA. This distorted phase exhibits two distinct Mo-S bond lengths (2.41~\AA{} and 4.50~\AA{}) and three types of Mo-Mo separations (2.75~\AA{}, 3.15~\AA{}, and 3.76~\AA{}), reflecting its reduced symmetry.
Energetically, 2H-MoS$_2$ is more stable, with the 1T$^\prime$ phase lying approximately 0.19~eV per atom higher in energy. Nonetheless, experimental studies show that 1T$^\prime$-MoS$_2$ can be stabilized via chemical doping, strain engineering, or electrochemical control,\cite{lukowski2013enhanced,gonzalez2024mos2}, making it a viable candidate for battery applications where metallic character is advantageous.

\begin{figure}[!]
\centering
\includegraphics[clip,width=0.9\columnwidth,angle=0]{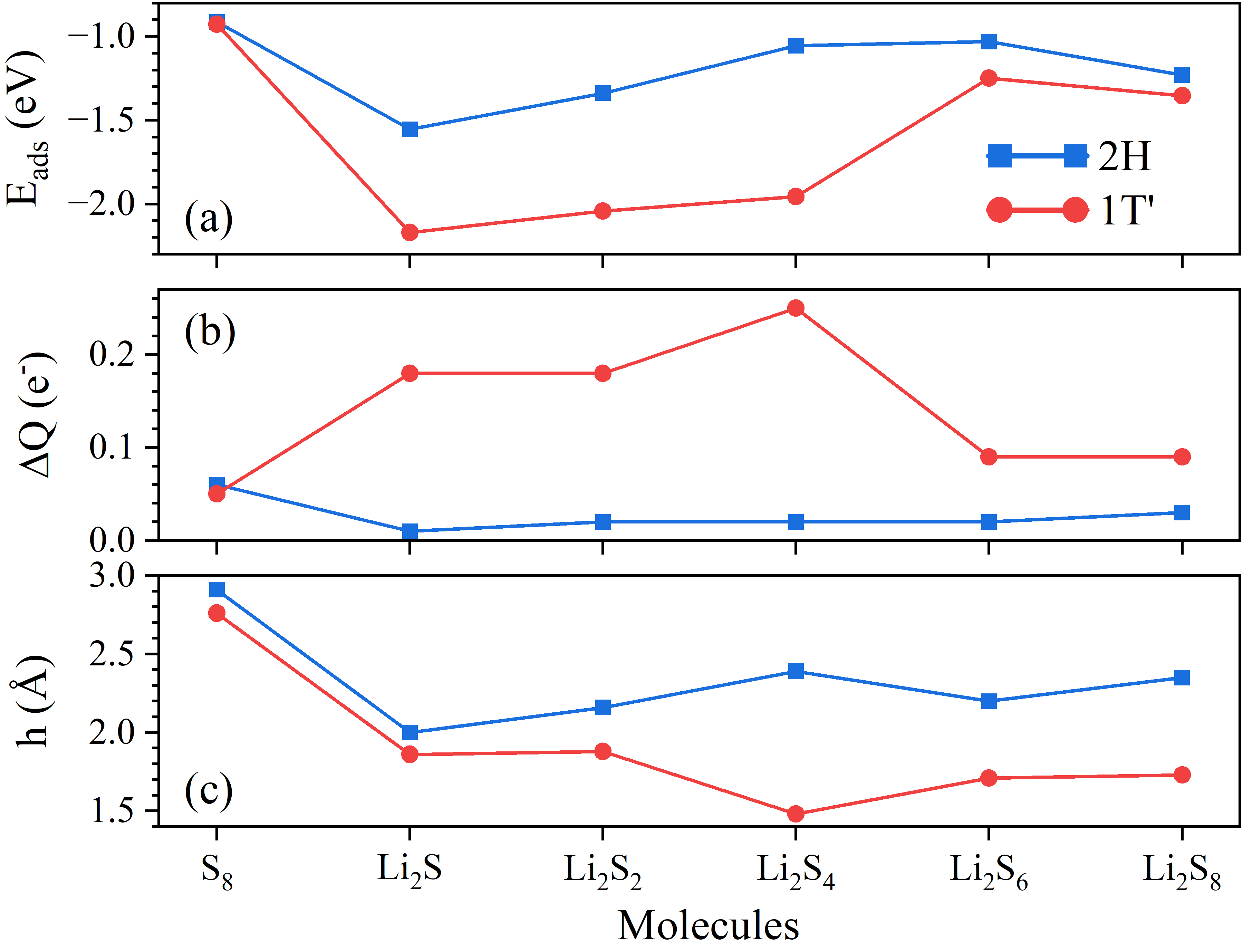} 
\caption{Li$_m$S$_n$ adsorption properties on MoS$_2$ monolayers. Panel (a) shows the adsorption energy ($E_{\text{ads}}$) for the 2H and 1T$^\prime$ phases, panel (b) shows the charge transfer ($\Delta Q$) from the Li$_m$S$_n$ molecules to the MoS$_2$ monolayer (positive values indicate charge transfer to MoS$_2$), and panel (c) illustrates the vertical distance ($h$) between the molecules and the surface. Blue squares represent the 2H-MoS$_2$ phase, and red circles represent the 1T$^\prime$-MoS$_2$ phase.
}
\label{Fig:LiSgeneral}
\end{figure}

\subsection{Adsorption Properties of Li$_m$S$_n$ Molecules\label{sec:bind}}
We evaluate the adsorption of Li$_m$S$_n$ molecules on both 2H and 1T$^\prime$ MoS$_2$ monolayers by examining three key descriptors: adsorption energy ($E_{\mathrm{ads}}$), charge transfer ($\Delta Q$), and vertical distance ($h$) between the adsorbate and the surface, as shown in Figure~\ref{Fig:LiSgeneral}. The adsorption energy is defined as follows:
\begin{equation}
E_{\mathrm{ads}} = E_{\mathrm{full}} - E_{\mathrm{Li}_m\mathrm{S}_n} - E_{\mathrm{MoS_2}},
\end{equation}
where $E_{\mathrm{full}}$ is the total energy of the Li$_m$S$_n$ molecule adsorbed on MoS$_2$, $E_{\mathrm{Li}_m\mathrm{S}_n}$ is the energy of the isolated molecule (computed in a $30 \times 30 \times 30$~\AA{} cell), and $E_{\mathrm{MoS_2}}$ is the energy of pristine MoS$_2$. A negative $E_{\mathrm{ads}}$ indicates an exothermic and thermodynamically favorable interaction.

Figure~\ref{Fig:LiSgeneral}(a) shows that 1T$^\prime$-MoS$_2$ exhibits stronger adsorption energies than 2H-MoS$_2$, especially for smaller polysulfides such as Li$_2$S and Li$_2$S$_2$. In the context of Li--S batteries, this suggests that 1T$^\prime$-MoS$_2$ more effectively anchors polysulfides, suppressing the shuttle effect and improving cycling stability and energy efficiency.\cite{manthiram2020reflection,su2012lithium}

We define the charge transfer $\Delta Q$ as:
\begin{equation}
\Delta Q = Q_{\mathrm{full}} - Q_{\mathrm{MoS_2}},
\end{equation}
where $Q_{\mathrm{full}}$ is the total charge of the MoS$_2$ monolayer with adsorbed Li$_m$S$_n$, and $Q_{\mathrm{MoS_2}}$ is the charge of the pristine monolayer. A positive $\Delta Q$ indicates electron donation from the adsorbate to the surface.
 Figure~\ref{Fig:LiSgeneral}(b) confirms this behavior across all configurations, with 1T$^\prime$-MoS$_2$ exhibiting higher $\Delta Q$ values than 2H-MoS$_2$, particularly for intermediate species such as Li$_2$S$_2$ and Li$_2$S$_4$. This enhanced charge transfer strengthens the binding interaction and further suppresses polysulfide diffusion.\cite{manthiram2020reflection}

\begin{figure}[!]
\centering
\includegraphics[clip,width=0.9\columnwidth,angle=0]{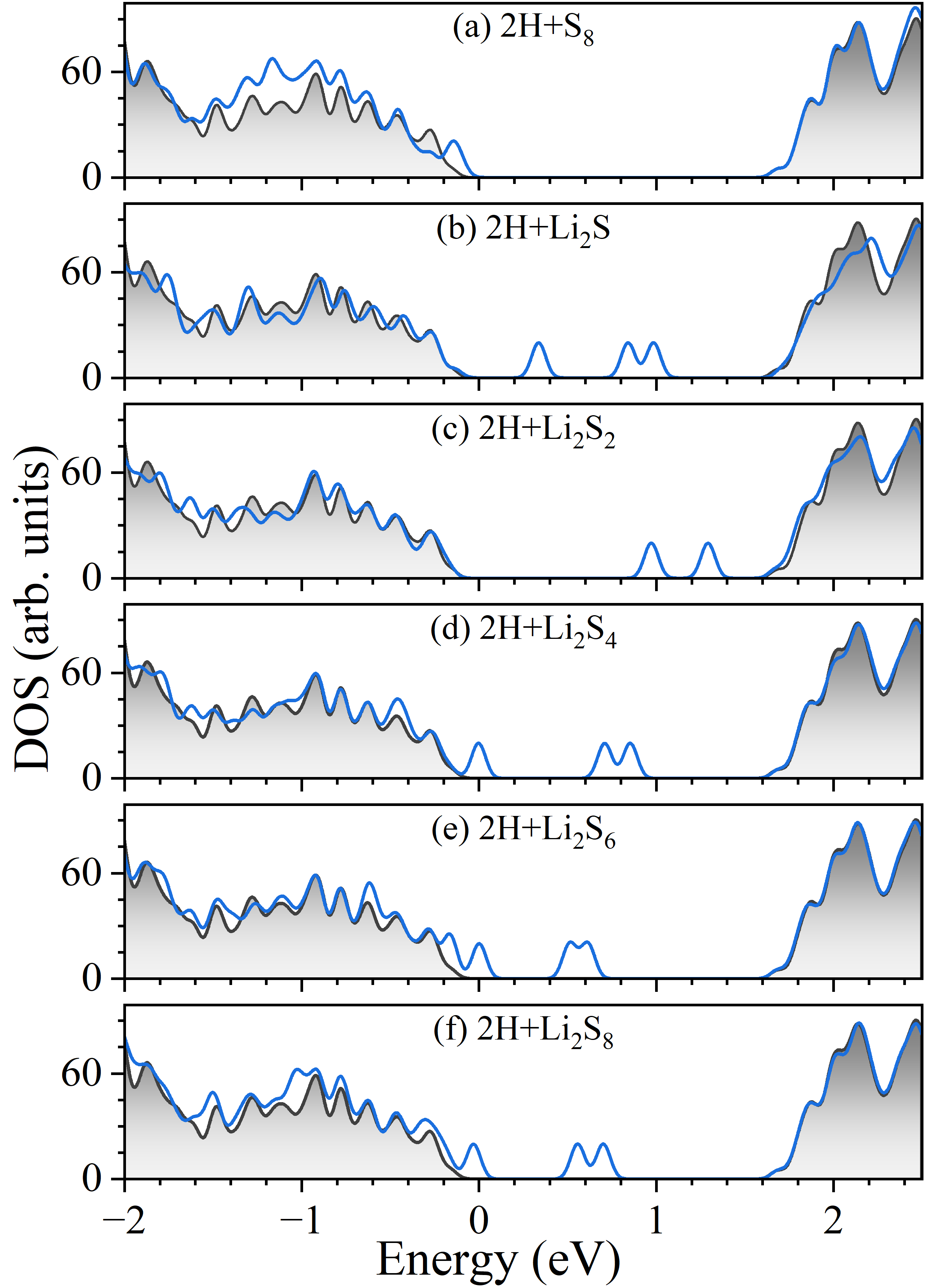} 
\caption{The density of states (DOS) of 2H-MoS$_2$ with adsorbed Li$_m$S$_n$ molecules, compared with the pristine 2H-MoS$_2$ DOS. Different panels correspond to the DOS for 2H-MoS$_2$ with Li$_m$S$_n$ molecules: S$_8$ in (a),  Li$_2$S in (b), Li$_2$S$_2$ in (c), Li$_2$S$_4$ in (d), Li$_2$S$_6$ in (e), and Li$_2$S$_8$ in (f), respectively. Blue lines represent the DOS of the 2H-MoS$_2$ system with an adsorbed molecule, while the pristine DOS is shown as a shaded background for comparison.
}
\label{Fig:2Hdos}
\end{figure}

\begin{figure}[h!]
\centering
\includegraphics[width=0.9\columnwidth]{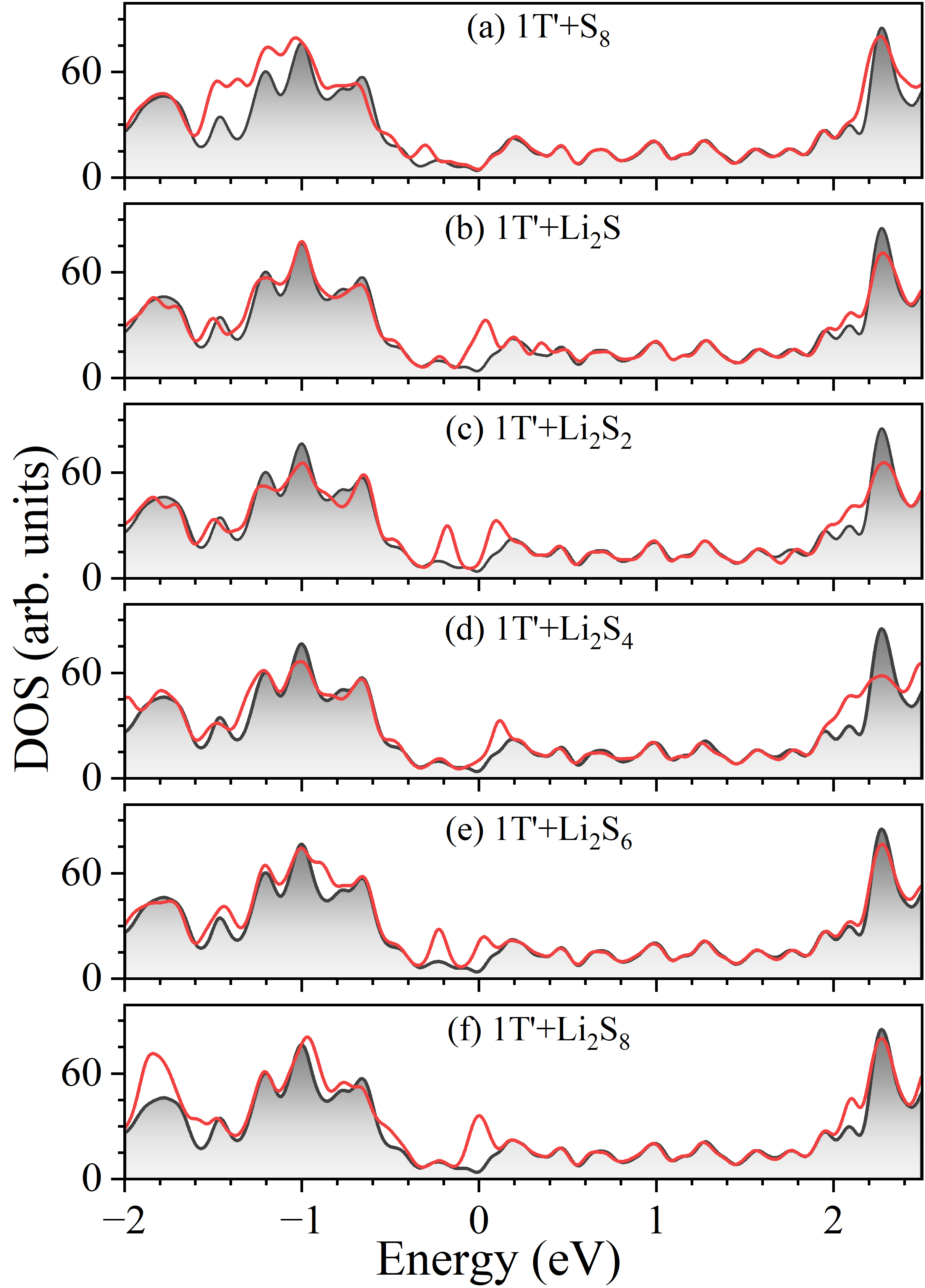} 
\caption{The density of states (DOS) of 1T$^\prime$-MoS$_2$ with adsorbed Li$_m$S$_n$ molecules, compared with the pristine 1T$^\prime$-MoS$_2$ DOS (shaded region). The panel distribution and labeling are the same  as in Fig.~\ref{Fig:2Hdos}.}
\label{Fig:T1pdos}
\end{figure}

Figure~\ref{Fig:LiSgeneral}(c) presents the vertical distance $h$ from the average topmost sulfur atoms of MoS$_2$ to the nearest atom in the Li$_m$S$_n$ cluster. The 1T$^\prime$ phase consistently yields shorter adsorption heights than the 2H phase, especially for Li$_2$S and Li$_2$S$_2$, reinforcing its role as a more effective anchoring substrate.\cite{su2012lithium}

Overall, the 1T$^\prime$-MoS$_2$ phase exhibits superior performance compared to 2H-MoS$_2$ in terms of adsorption strength, charge donation, and interfacial proximity, indicating a greater ability to anchor polysulfides and facilitate redox reactions. This behavior is consistent with prior studies reporting that metallic TMDs generally interact more strongly with polysulfide species than their semiconducting counterparts.\cite{hojaji2022dft,liu2021situ} However, this enhanced interaction can also pose potential limitations: overly strong binding may hinder the desorption of reaction intermediates, particularly during the late stages of the redox cycle, and reduce reversibility under dynamic conditions.\cite{huang2018first,hojaji2022dft}  These findings suggest that while 1T$^\prime$ domains offer significant advantages for polysulfide immobilization, careful control of their coverage and local environment is necessary to avoid kinetic bottlenecks associated with overbinding.
 Nevertheless, the weaker interaction of 2H-MoS$_2$ may offer advantages by enabling controlled polysulfide mobility, potentially facilitating redox reversibility while limiting over-anchoring. This tradeoff highlights the value of hybrid systems that combine both phases to optimize sulfur cathode performance in Li--S batteries.

Our DFT results are in agreement with previous experimental reports. For instance, oxygen-doped MoS$_2$ with partial 1T character achieves a discharge capacity of 1015~mAh\,g$^{-1}$ with 78.5\% retention over 300 cycles,\cite{ren2024impact} consistent with enhanced charge transfer in metallic domains. Similarly, 1T-MoS$_2$/MnO$_2$ composites promote a transition from liquid-solid to solid-solid Li$_2$S formation, enabling more uniform nucleation.\cite{Qing2024} Additionally, MCNT/MoS$_2$/S hybrids maintain capacities near 460~mAh\,g$^{-1}$ after 300 cycles,\cite{Guo2020} although the influence of conductive additives complicates direct interpretation.

\subsection{\label{sec:elect}Electronic Properties}
Figures~\ref{Fig:2Hdos} and \ref{Fig:T1pdos} present the density of states (DOS) for 2H- and 1T$^\prime$-MoS$_2$ monolayers, respectively, following the adsorption of various Li$_m$S$_n$ species. For comparison, each panel also displays the DOS of the pristine MoS$_2$ monolayer (shaded area), allowing for a direct assessment of the electronic modifications induced by polysulfide adsorption.

In 2H-MoS$_2$, the adsorption of S$_8$ and Li$_m$S$_n$ species introduces localized states near the valence band maximum (VBM), as seen in Figures~\ref{Fig:2Hdos}(a)-(e). For example, Li$_2$S$_4$ and Li$_2$S$_6$ produce gap states just above the VBM, indicating mild hybridization and limited charge transfer.\cite{gonzalez2024mos2} Notably, the band gap remains largely preserved after adsorption, reflecting the weak interaction and localized nature of these induced states. Similar behavior has been reported in other semiconducting TMDs, such as WS$_2$ and MoSe$_2$, where the intrinsic gap is retained despite adsorbate-induced features.\cite{su2012lithium} From a device standpoint, preserving the band gap can help limit parasitic conduction and reduce polysulfide shuttling.

By contrast, 1T$^\prime$-MoS$_2$  phase exhibits significant changes in its DOS upon adsorption. Even for S$_8$ [Figure~\ref{Fig:T1pdos}(a)], significant hybridization occurs, with enhanced density of states appearing below the Fermi level (E$_F$). This effect intensifies with lithium-containing species: Li$_2$S$_4$ and Li$_2$S$_6$ induce substantial DOS redistribution near E$_F$ [Figures~\ref{Fig:T1pdos}(d) and (e)], consistent with the stronger charge transfer and binding energies reported in Section~\ref{sec:bind}. These changes reflect the metallic character of 1T$^\prime$-MoS$_2$ and its ability to accept electrons from Li$_m$S$_n$, facilitating improved electronic coupling and redox activity.
These distinct responses arise from the inherent electronic differences between the two polymorphs. The semiconducting 2H-MoS$_2$ shows minimal perturbation upon adsorption, with only weakly localized states near the band edges. In contrast, the metallic 1T$^\prime$ phase undergoes substantial electronic restructuring, indicating stronger adsorbate-substrate interactions. These findings agree with previous studies  that report improved electron accommodation  and catalytic activity in metallic TMDs.\cite{manthiram2020reflection}

From a practical perspective, combining both polymorphs offers a promising strategy: 1T$^\prime$-MoS$_2$ provides high conductivity and abundant active sites for redox reactions, while 2H-MoS$_2$ may help stabilize the structure and suppress undesired electron leakage. This concept of phase complementarity is increasingly adopted in Li--S cathode design, leveraging the synergy between conductive and semiconducting domains to achieve both efficient charge transport and effective sulfur confinement.\cite{lukowski2013enhanced}

\begin{figure}[!]
\centering
\includegraphics[clip,width=0.9\columnwidth,angle=0]{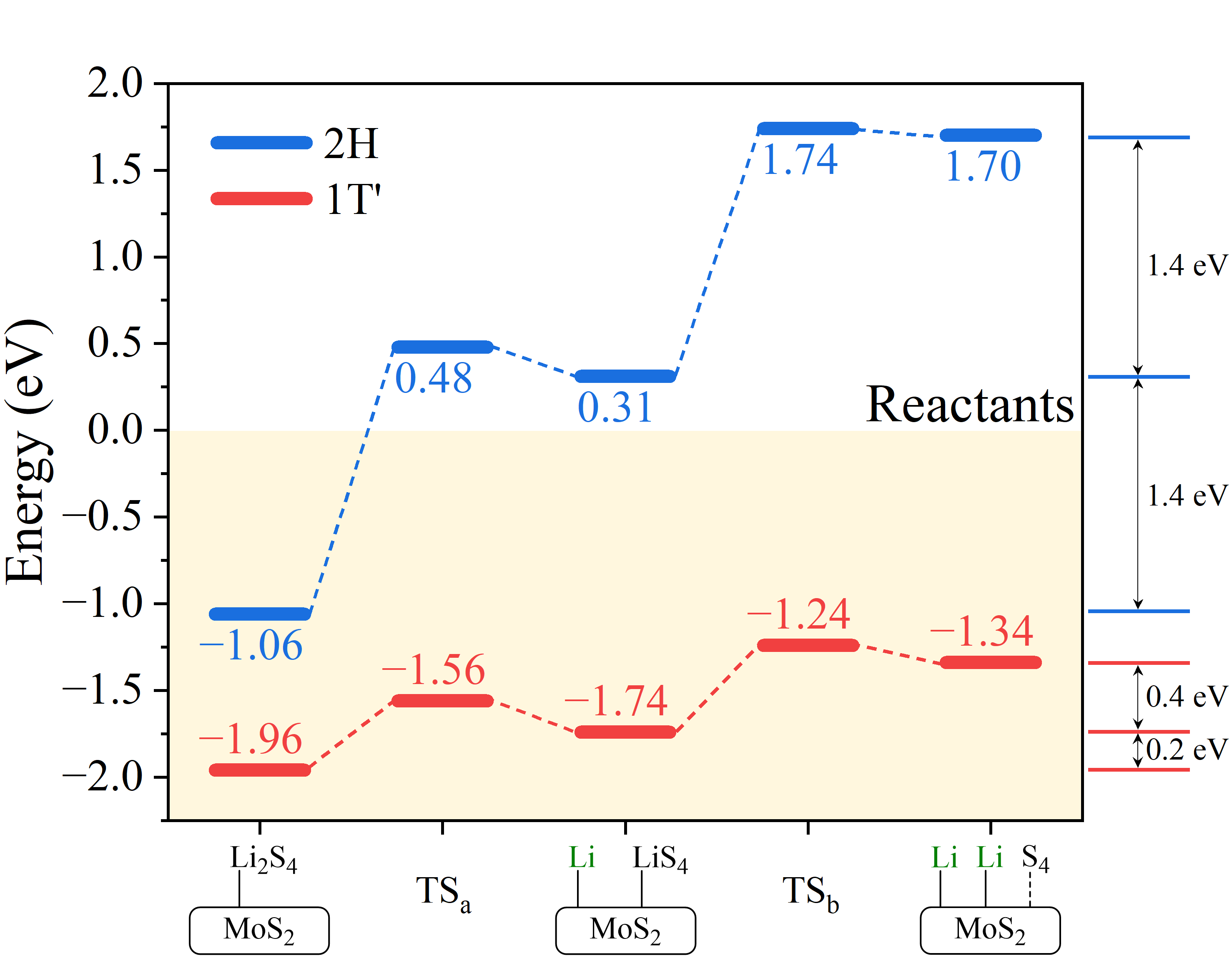} 
\caption{NEB calculation for the Li$_2$S$_4$ delithiation process on 2H- and 1T$^\prime$-MoS$_2$ surfaces. The external markers indicate the energies of metastable configurations along the reaction pathway. Blue lines represent the 2H phase, and red lines represent the 1T$^\prime$ phase. Energies are given in electron volts (eV) relative to the reactants (energy of isolated molecule plus pristine surface).
}
\label{Fig:neb}
\end{figure}

\begin{figure*}[!]
\centering
\includegraphics[clip,width=0.8\textwidth,angle=0]{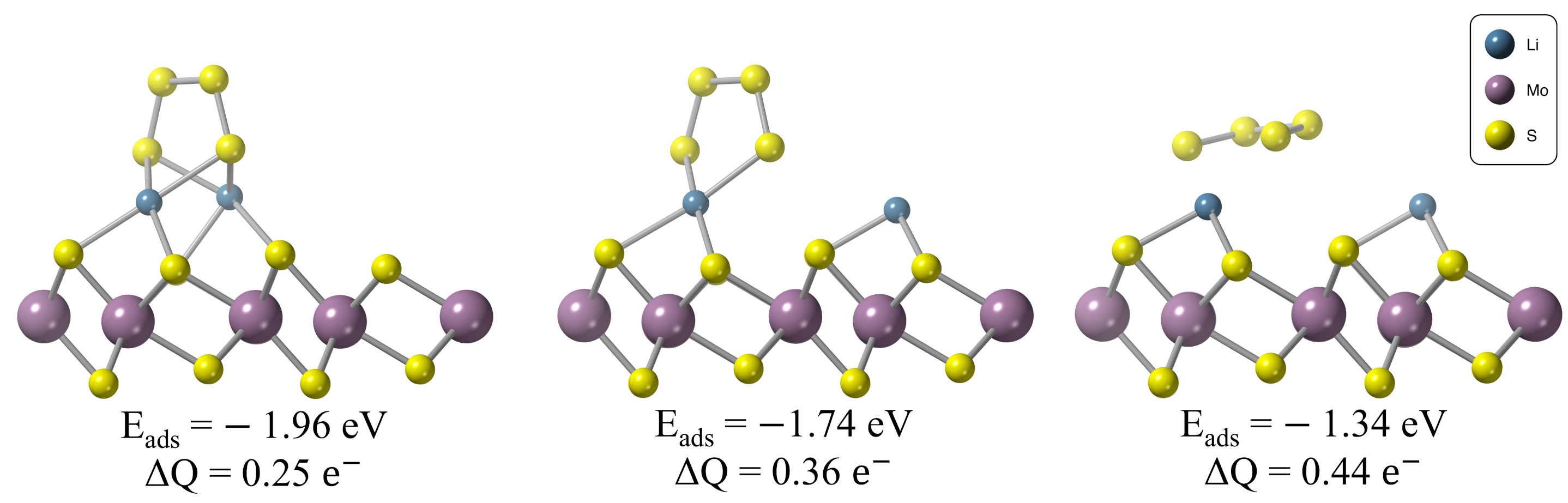} 
\caption{Li$_2$S$_4$ delithiation process on the 1T$^\prime$-MoS$_2$ surface. The adsorption energies ($E_{\text{ads}}$) for each step are shown below the structures. The color scheme follows Fig.~\ref{Fig:scheme}. 
}
\label{Fig:deli_1T}
\end{figure*}

\subsection{\label{sec:neb}Delithiation Pathway}

Figure~\ref{Fig:neb} displays the minimum energy paths for the delithiation of Li$_2$S$_4$ adsorbed on 2H- and 1T$^\prime$-MoS$_2$, computed using the nudged elastic band (NEB) method. In both cases, the initial adsorption of Li$_2$S$_4$ is exothermic, with the 1T$^\prime$ surface exhibiting stronger binding ($E_{\mathrm{ads}} = -1.96$\,eV) than the 2H phase ($E_{\mathrm{ads}} = -1.06$\,eV). This enhanced interaction reflects the higher affinity of metallic transition metal dichalcogenides (TMDs) for polar sulfur species, consistent with prior experimental and theoretical reports.\cite{manthiram2020reflection}
To gain mechanistic insight into the ion release process, we analyze the delithiation pathway in two successive steps: the removal of the first Li$^+$ ion (transition state $\text{TS}_a$) and the subsequent detachment of the second Li$^+$ ion (transition state $\text{TS}_b$). The corresponding energy barriers and intermediate stabilities reveal sharp contrasts in the kinetics and thermodynamics of the two MoS$_2$ polymorphs.

\paragraph*{First Delithiation Step ($\text{TS}_a$):}
The initial delithiation step proceeds as $
\text{Li}_2\text{S}_4 \rightarrow \text{LiS}_4 + \text{Li}^+.
$
This reaction is kinetically hindered on the semiconducting 2H surface, with an activation energy of 1.54\,eV and the LiS$_4$ intermediate positioned 1.40\,eV above the initial state. Such a high barrier suggests limited ion mobility and sluggish redox kinetics under ambient conditions. In contrast, the same process on the 1T$^\prime$ phase features a much lower barrier of 0.40\,eV, and the intermediate lies only 0.20\,eV above the starting configuration. These values indicate that Li$^+$ detachment is significantly more accessible on 1T$^\prime$-MoS$_2$, consistent with its superior electronic and ionic transport properties.\cite{manthiram2020reflection}

\paragraph*{Second Delithiation Step ($\text{TS}_b$):}
The subsequent step releases the remaining lithium ion:
$
    \text{LiS}_4 \rightarrow \text{S}_4 + 2\text{Li}^+.
$
Once again, the 1T$^\prime$ surface enables a favorable pathway, with a barrier of 0.40\,eV and a final S$_4$ product that is thermodynamically stable ($E = -1.34$\,eV relative to the initial Li$_2$S$_4$ state). In contrast, the same transformation on 2H-MoS$_2$ is much less favorable, requiring an activation energy of 1.43\,eV and producing an unstable S$_4$ species 1.70\,eV above the initial state. These results emphasize the dual kinetic and thermodynamic limitations of the 2H phase in enabling efficient polysulfide conversion.

%%\paragraph*{Reaction Landscape and Intermediate Stabilization:}

Figure~\ref{Fig:deli_1T} complements the NEB results by depicting the fully relaxed atomic configurations and corresponding adsorption energies along the delithiation sequence on 1T$^\prime$-MoS$_2$. Throughout the process-from Li$_2$S$_4$ to LiS$_4$ to S$_4$-the system maintains strong interfacial coupling, accompanied by a progressive increase in charge transfer ($\Delta Q$). Specifically, $\Delta Q$ increases from 0.25\,e$^-$ in the initial state to 0.36\,e$^-$ for LiS$_4$, and further to 0.44\,e$^-$ for the final S$_4$ product. This monotonic growth in $\Delta Q$ reflects efficient charge accommodation by the metallic substrate, enabling stabilization of partially oxidized intermediates via delocalized electronic states.
The smooth energy landscape and persistent binding interactions suggest a continuous and energetically favorable reaction path that supports rapid redox kinetics and effective sulfur utilization. These properties, combined with reduced reaction barriers, reinforce the role of 1T$^\prime$-MoS$_2$ as a highly effective anchoring and catalytic platform for lithium polysulfide conversion in Li--S batteries.\cite{manthiram2020reflection}

%\paragraph*{Metastable States and Kinetic Favorability:}
The energy landscape of the delithiation process differs markedly between the two MoS$_2$ polymorphs. The 1T$^\prime$ phase exhibits low and relatively uniform barriers between successive intermediates, indicative of a smooth reaction pathway with minimal structural rearrangement. In contrast, the 2H phase shows abrupt energy variations between states, indicating a slower and energetically demanding process.

Figure~\ref{Fig:deli_1T} illustrates the optimized structures and key electronic descriptors along the delithiation sequence on 1T$^\prime$-MoS$_2$. The initial adsorption of Li$_2$S$_4$ is strongly exothermic ($E_{\mathrm{ads}} = -1.96$\,eV), accompanied by a charge transfer of $\Delta Q = 0.25\,e^-$. Upon removal of the first Li$^+$ ion, the resulting LiS$_4$ intermediate remains strongly bound ($E_{\mathrm{ads}} = -1.74$\,eV), with an increased charge transfer of $\Delta Q = 0.36\,e^-$. Following the second delithiation, the final S$_4$ species also remains stable ($E_{\mathrm{ads}} = -1.34$\,eV), while $\Delta Q$ rises to 0.44\,e$^-$.
The progressive increase in $\Delta Q$ along the reaction coordinate reflects the ability of 1T$^\prime$-MoS$_2$ to redistribute and accommodate charge efficiently, stabilizing partially oxidized intermediates. As reported in related metallic TMD systems,\cite{lukowski2013enhanced} such charge delocalization correlates with enhanced redox kinetics and higher rate capability.

%These kinetic insights highlight the favorable energy landscape of the 1T$^\prime$ phase, but a complete picture of polysulfide reactivity must also consider temperature effects and solvent interactions. We therefore proceed to examine the thermodynamic stability of these species under operating conditions.

%%%% NEW
\subsection{Comparison 1T-MoS$_2$ Polymorph}

The surface reactivity of MoS$_2$ polymorphs varies significantly with their electronic structure and local coordination. Our DFT results show that the 2H-MoS$_2$ phase exhibits moderate polysulfide binding, with adsorption energies ranging from $-1.10$ to $-0.90$\,eV across the Li$_m$S$_n$ series. These values may be insufficient to fully suppress polysulfide diffusion and ensure robust redox mediation, limiting the catalytic efficiency of this phase.
In contrast, the 1T$^\prime$-MoS$_2$ phase exhibits substantially stronger interactions, with adsorption energies as low as $-1.96$\,eV for Li$_2$S$_4$ and $-1.34$\,eV for the final S$_4$ product. The corresponding activation barriers for delithiation (0.40\,eV for both Li$_2$S$_4 \rightarrow$ LiS$_4$ and LiS$_4 \rightarrow$ S$_4$) are significantly lower than those of the 2H phase (1.54 and 1.43\,eV), indicating more favorable reaction kinetics under operating conditions~\cite{gonzalez2024mos2}.
These trends are consistent with the behavior reported for the pristine and defect-engineered 1T-MoS$_2$ phase. Previous reports\cite{hojaji2022dft}, the adsorption energies for Li$_2$S, Li$_2$S$_2$, and Li$_2$S$_4$ on pristine 1T-MoS$_2$ lie between $0.7$–$3.3$\,eV, while defected surfaces with one or two sulfur vacancies reach values as high as 5.38\,eV for Li$_2$S and 4.04\,eV for Li$_2$S$_4$~\cite{hojaji2022dft}. While such strong interactions enhance polysulfide retention, they may hinder desorption and redox reversibility. Indeed, the activation barrier for the Li$_2$S$_2 \rightarrow$ Li$_2$S reaction reaches 2.30\,eV on the 1T-2S surface, compared to just 0.40\,eV on 1T$^\prime$-MoS$_2$~\cite{hojaji2022dft,zhang2019sulfur}.

Therefore, 1T$^\prime$-MoS$_2$ structure offers a more balanced profile: it exhibits sufficient binding to suppress polysulfide migration while maintaining low kinetic barriers to sustain redox activity. This intermediate behavior avoids the desorption bottlenecks of overbinding observed in vacancy-rich 1T phases, suggesting that 1T$^\prime$ may serve as a more versatile and tunable platform for Li–S battery cathodes.

%%%%%

\subsection{Thermochemistry (Gas-phase approach)}

To evaluate the thermodynamic viability of polysulfide adsorption, we compute Gibbs free energy differences, $\Delta G$, using a gas-phase cluster model. This approach enables the inclusion of enthalpic and entropic contributions without relying on full periodic vibrational calculations, which remain computationally demanding, particularly when combined with solvation models.
This method enables efficient and internally consistent evaluation of vibrational properties and thermal corrections, yielding accurate free energy estimates at finite temperature while avoiding the complexity and potential artifacts of periodic phonon calculations under solvation. Moreover, cluster-based approaches have been shown to produce results in close agreement with periodic calculations for systems where adsorption effects are spatially localized, further validating their use in this context.

At the time of writing, solvation models are not natively implemented in VASP for vibrational analyses. While alternative schemes such as VASPsol are available, we adopt a quantum chemistry-based strategy successfully applied in prior studies.\cite{reuter2012first,kurfman2021calculating,scott2012predictions} Specifically, we extract a representative molecular cluster from the 1T$^\prime$-MoS$_2$ surface and place the Li$_m$S$_n$ species at its geometric center. The cluster is sufficiently large to suppress edge effects and retain the local structural and electronic environment of the periodic system.

%\subsubsection*{Free Energy Computation and Solvation}

The thermodynamic stability of adsorbed polysulfides depends not only on their binding strength at 0\,K, but also on the temperature and the chemical environment-particularly the presence of solvent. These factors are critical for real operating conditions in Li--S batteries, where temperature variations and electrolyte composition influence desorption, shuttling, and redox reversibility. To capture these effects, we compute Gibbs free energy changes ($\Delta G$) for Li$_m$S$_n$ adsorption on 1T$^\prime$-MoS$_2$ across a range of temperatures and under both gas-phase and solvated conditions.
Within this framework, we compute the Gibbs free energy change for adsorption as:
\begin{equation}
    \Delta G = G_{\mathrm{full}}(P,T) - G_{\mathrm{Li}_m\mathrm{S}_n}(P,T) - G_{\mathrm{MoS_2}}(P,T),
\end{equation}
where each term corresponds to the free energy of the adsorbate-cluster complex, the isolated Li$_m$S$_n$ molecule, and the pristine MoS$_2$ cluster, respectively.

We include solvation effects using the conductor-like polarizable continuum model (CPCM) to account for the electrolyte environment in operating Li–S batteries.\cite{Tomasi2005,barone1998quantum} As a representative medium, we adopt dimethyl carbonate (DMC), a widely used solvent in lithium-based electrolytes, characterized by a dielectric constant of $\varepsilon = 3.12$ and a refractive index of $n_D = 1.419$.\cite{Xu2004} This implicit solvation model captures the dielectric screening provided by the electrolyte, stabilizing polar or charged species and modulating the entropic contribution to adsorption, thereby influencing the overall free energy landscape.

\begin{table}[h]
    \centering
    \begin{tabular}{c|c|c}
        \hline\hline
        Temperature & \textbf{Without} solvation & \textbf{With} solvation\\
        (K)         & $\Delta G$ (eV)           & $\Delta G$ (eV)     \\
        \hline
        253.15 & -0.461 & -0.311 \\
        273.15 & -0.419 & -0.267 \\
        293.15 & -0.377 & -0.222 \\
        313.15 & -0.335 & -0.178 \\
        \hline\hline
    \end{tabular}
    \caption{Gibbs free energy change ($\Delta G$) for Li$_m$S$_n$ adsorption on 1T$^\prime$-MoS$_2$ at different temperatures (1\,atm), comparing gas-phase and solvated (CPCM-DMC) models.}
    \label{tab:Gibbs}
\end{table}

Table~\ref{tab:Gibbs} summarizes the computed Gibbs free energy changes ($\Delta G$) for Li$_m$S$_n$ adsorption on 1T$^\prime$-MoS$_2$ across a range of temperatures, comparing gas-phase and solvated conditions. Adsorption remains thermodynamically favorable ($\Delta G < 0$) in all cases; however, the presence of solvation consistently reduces the magnitude of $\Delta G$, a trend that becomes more pronounced at higher temperatures due to the increasing entropic cost of adsorption.

As shown in Table~\ref{tab:Gibbs}, implicit solvation leads to systematically less negative $\Delta G$ values, indicating that solvent screening weakens the adsorbate–surface interaction. This behavior aligns with prior reports showing that polar environments reduce electrostatic binding between charged species and surfaces.\cite{Andreussi2012,Tomasi2005}
For instance, at 293\,K, near typical battery operating temperatures, the free energy of adsorption decreases from $-0.377$\,eV in the gas phase to $-0.222$\,eV under solvation, representing a reduction of approximately 41\%. This difference becomes even more significant at 313\,K, where $\Delta G$ drops from $-0.335$\,eV (gas) to $-0.178$\,eV (solvated), reflecting the growing impact of entropic penalties at elevated temperatures.

From a design perspective, electrolyte formulating electrolytes requires balancing two competing effects: interactions strong enough to immobilize polysulfides and suppress the shuttle effect yet sufficiently weak to avoid overbinding that could hinder redox kinetics. Although DMC solvation reduces adsorption strength, it may enhance molecular mobility and help prevent electrode passivation.
Solvation also affects the temperature threshold at which adsorption becomes thermodynamically unfavorable. Our calculations place this crossover near 473\,K in the gas phase and approximately 393\,K under implicit solvation, showing the role of the solvent in tuning the thermal stability and operational regimes of adsorbed polysulfide species.\cite{manthiram2020reflection,su2012lithium}

\begin{figure}[!]
\centering
\includegraphics[clip,width=0.95\columnwidth,angle=0]{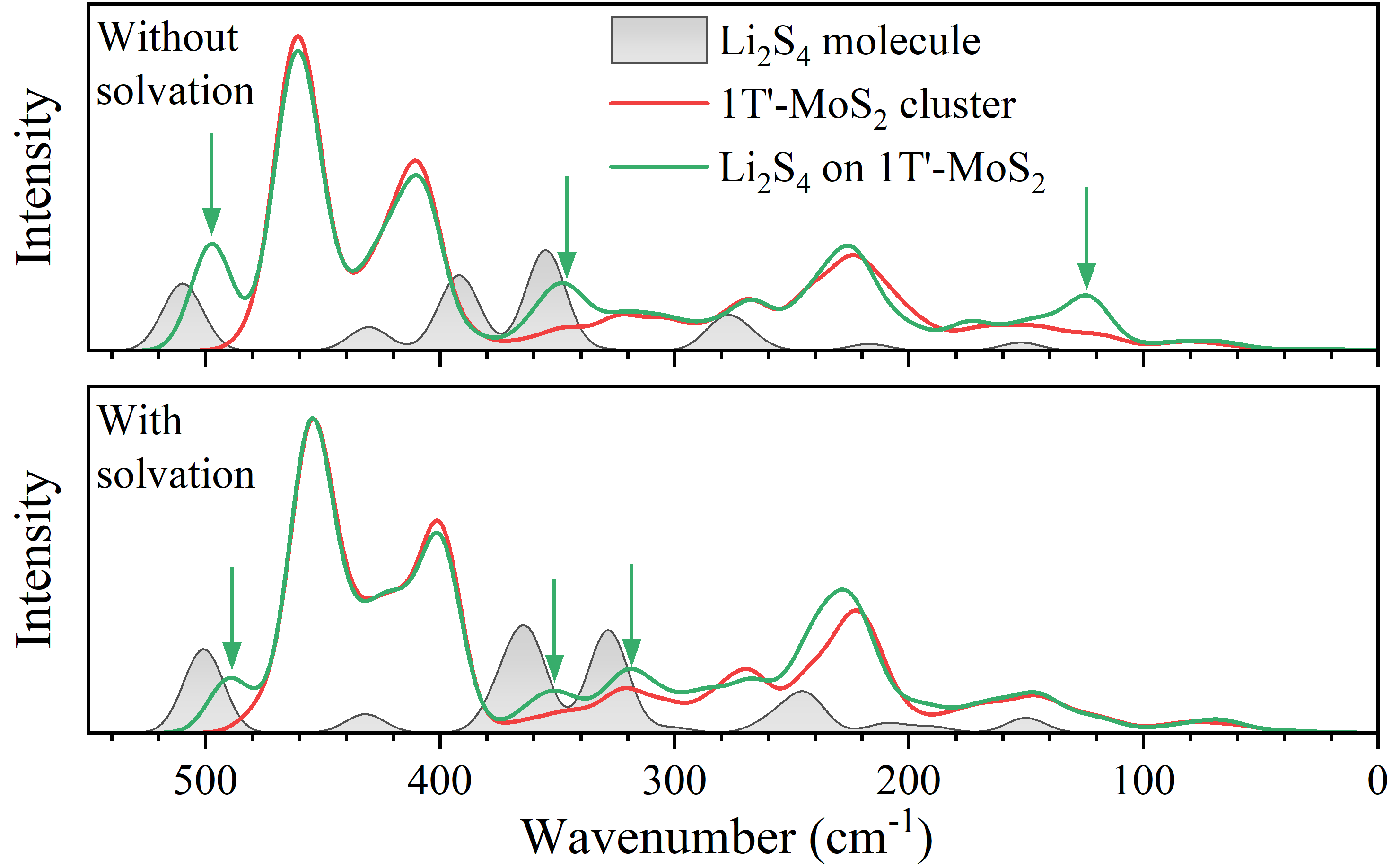} 
\caption{Infrared (IR) spectra of Li$_2$S$_4$, 1T$^\prime$-MoS$_2$, and Li$_2$S$_4$ adsorbed on 1T$^\prime$-MoS$_2$, calculated using ORCA-DFT. The upper panel corresponds to gas-phase calculations without solvation effects, while the lower panel includes solvation effects. The arrows indicate the main vibrational features induced by the Li$_2$S$_4$ molecule on the 1T$^\prime$-MoS$_2$ surface.
}
\label{fig:IR_spectra}
\end{figure}

\subsection{Spectral Analysis}
Infrared (IR) spectroscopy serves as a direct and experimentally accessible probe of molecular bonding environments and interfacial interactions. To complement the thermodynamic analysis, Figure~\ref{fig:IR_spectra} displays the IR spectra of Li$_2$S$_4$ adsorbed on a 1T$^\prime$-MoS$_2$ cluster, computed under both gas-phase and solvated conditions. The spectra, obtained from DFT calculations using ORCA, exhibit pronounced vibrational shifts upon adsorption, reflecting strong chemical coupling between the polysulfide species and the metallic surface.

In the gas phase, Li$_2$S$_4$ exhibits pronounced absorption features between 100 and 500~cm$^{-1}$, primarily attributed to Li–S stretching and bending modes. Compared to the isolated molecule, these modes experience moderate redshifts (10–20~cm$^{-1}$), indicating bond softening and geometric rearrangement upon surface binding. This behavior agrees with previous findings that TMD substrates can modulate vibrational force constants and induce conformational distortions in adsorbed polysulfides.\cite{su2012lithium}

Further spectral modifications are observed when solvation is included via the CPCM model, particularly at lower wavenumbers. Both peak positions and intensities change, suggesting that the dielectric environment influences local electric fields and vibrational coupling. These effects are consistent with the thermodynamic trend of weakened adsorption in solution and indicate notable electronic and structural reorganization at the solid-liquid interface.\cite{Tomasi2005,barone1998quantum,Xu2004}

Significantly, adsorption on 1T$^\prime$-MoS$_2$ consistently induces downshifts in several Li–S modes, often associated with bond elongation or weakening due to interfacial charge transfer and substrate-induced strain.\cite{Zhao2008} These vibrational signatures correlate with the increase in charge transfer ($\Delta Q$) and stronger adsorption energies ($E_{\mathrm{ads}}$), reinforcing the role of 1T$^\prime$-MoS$_2$ as an electronically responsive and structurally adaptive host for polysulfide immobilization.
Importantly, the predicted spectral features offer clear experimental targets for verification. Infrared and Raman spectroscopy can directly probe the Li--S stretching and bending modes and monitor their redshifts upon adsorption on MoS$_2$ surfaces. By tracking the emergence of new vibrational features associated with Mo-S-Li interfacial interactions and following the evolution of these peaks during battery cycling, experimental studies can sensitively assess the anchoring behavior, catalytic activity, and phase identity predicted by our calculations.

\section{Final Remarks}

This study employs first-principles calculations to evaluate the potential of two MoS$_2$ polymorphs, 2H and 1T$^\prime$, as anchoring materials for lithium-sulfur (Li--S) battery cathodes. We assess their ability to suppress polysulfide shuttling and enhance redox kinetics through detailed analysis of their electronic structure, adsorption energetics, and thermochemical responses.
Our results demonstrate that the metallic 1T$^\prime$ phase consistently outperforms the semiconducting 2H phase in adsorption strength, charge transfer, and electronic coupling. These properties enable stronger immobilization of lithium polysulfides and facilitate faster redox dynamics. Conversely, the 2H phase, despite its weaker interactions, may improve mechanical integrity and reduce parasitic conductivity, suggesting a complementary role in hybrid architectures.
Thermodynamic calculations reveal that solvation substantially weakens adsorption, lowering the polysulfide desorption threshold by approximately 80 K. This trend underscores the importance of considering solvent effects when designing interfacial chemistries. Simulated infrared spectra further confirm strong chemical coupling between Li$_2$S$_4$ and the 1T$^\prime$ surface, manifested as pronounced redshifts in vibrational modes, consistent with significant charge transfer and interfacial restructuring.
These predicted vibrational signatures provide clear experimental targets: infrared and Raman spectroscopy can directly probe the Li--S stretching and bending modes, track their redshifts upon adsorption, and monitor the emergence of new Mo--S--Li interfacial features during battery cycling, offering a sensitive route to validate the proposed anchoring and catalytic behavior.

From a materials design perspective, our results suggest that optimizing the coverage of 1T$^\prime$ domains can enhance polysulfide anchoring, but excessive metallic character may lead to overbinding and hinder desorption\cite{huang2018first,hojaji2022dft}. Operational stability may be improved by maintaining temperatures below the calculated desorption crossover (393 K under solvation), while targeted phase or defect engineering, through doping, strain, or electrochemical control, can be employed to tune the 2H/1T$^\prime$ ratio and activate interphase regions.
Specifically, chemical exfoliation, alkali metal intercalation, electron beam irradiation, and lithium intercalation under controlled potentials offer promising strategies to trigger the 2H to 1T$^\prime$ transition, providing tunable pathways to adjust the phase composition and optimize the balance between conductivity, stability, and catalytic performance. 
Additionally, electrolyte formulation should reduce polysulfide solubility while preserving sufficient mobility to avoid passivation or irreversible binding.

%% Experimental validation remains essential. In-situ techniques such as X-ray diffraction, Raman spectroscopy, and electrochemical impedance spectroscopy can monitor phase evolution and charge-transfer dynamics. In-situ IR and Raman measurements, in particular, could provide direct fingerprints of polysulfide-substrate interactions and validate vibrational shifts predicted by our simulations.

Beyond the specific case of Li--S batteries, the phase engineering principles identified here are broadly applicable to other alkali-metal-chalcogen battery systems, such as Na--S, K--S, and Li--Se chemistries. These systems share similar challenges related to shuttle effects, low electronic conductivity, and sluggish redox kinetics. Controlling the relative abundance of 2H and 1T$^\prime$ phases in MoS$_2$ or analogous two-dimensional materials offers a general strategy to tune interfacial interactions, enhance catalytic activity, and improve electrochemical performance across a wide range of emerging energy storage technologies. This phase-tuning approach could be further extended to hybrid solid-state architectures and multifunctional electrode designs, opening avenues for cross-platform improvements beyond the Li--S paradigm.

This work highlights the promise of phase-engineered MoS$_2$ as a modular platform for next-generation lithium-polysulfide cathodes. A synergistic approach integrating computational screening with targeted experimental efforts will be key to realizing durable, high-performance, and tunable energy storage systems. Taken together, these insights lay a solid foundation for the rational design of scalable, high-efficiency battery architectures and position phase-engineered two-dimensional materials as essential building blocks for advancing the frontiers of energy storage technology.

\section*{Acknowledgments}
JWG acknowledges financial support from ANID-FONDECY (Chile) grants N. 1220700 and 1221301.
RAG acknowledges financial support from Basal Program for Centers of Excellence, Grant AFB220001 (CEDENNA)

\section*{Competing Interests}
The Authors declare no Competing Financial or Non-Financial Interests.

\section*{Data Availability}
The data that support the findings of this study are available from the corresponding author, upon reasonable request.

\end{document}